# Phonons and Lithium diffusion in LiAlO$_2$

Mayanak K. Gupta[1,2$], Ranjan Mittal[1,3*], Baltej Singh[1,3], Olivier Delaire[2],
Srungarpu N. Achary[3,4], Stephane Rols[5], Avesh K. Tyagi[3,4] and Samrath L. Chaplot[1,3]

[1]*Solid State Physics Division, Bhabha Atomic Research Centre, Mumbai, 400085, India*
[2]*Department of Mechanical Engineering and Materials Science, Duke University, Durham NC, USA*
[3]*Homi Bhabha National Institute, Anushaktinagar, Mumbai 400094, India*
[4]*Chemistry Division, Bhabha Atomic Research Centre, Mumbai, 400085, India*
[5]*Institut Laue-Langevin, BP 156, 38042 Grenoble Cedex 9, France*
Email: mayankg@barc.gov.in[$], rmittal@barc.gov.in[*]

We report on investigations of phonons and lithium diffusion in LiAlO$_2$ based on inelastic neutron scattering (INS) measurements of the phonon density of states (DOS) in γ-LiAlO$_2$ from 473 K to 1073 K, complemented with ab-initio molecular dynamics (AIMD) simulations. We find that phonon modes related to Li vibrations broaden on warming as reflected in the measured phonon DOS and reproduced in simulations. Further, the AIMD simulations probe the nature of lithium diffusion in the perfect crystalline phase (γ-LiAlO$_2$), as well as in a structure with lithium vacancies and a related amorphous phase. Almost liquid-like super-ionic diffusion is observed in AIMD simulations of the three structures at high temperatures; with predicted onset temperatures of 1800 K, 1200 K, and 600 K in the perfect structure, vacancy structure and the amorphous phase, respectively. In the ideal structure, the Li atoms show correlated jumps; while simple and correlated jumps are both seen in the vacancy structure, and a mix of jumps and continuous diffusion occur in the amorphous structure. Further, we find that the Li-diffusion is favored in all cases by a large librational amplitude of the neighbouring AlO$_4$ tetrahedra, and that the amorphous structure opens additional diffusion pathways due to a broad distribution of AlO$_4$ tetrahedra orientations.

## I. INTRODUCTION

The study of mechanisms underlying Li diffusivity in Li-based compounds is of fundamental interest in research on Li-ion batteries and solid oxide fuel cells[1-11]. The stability and performance of a Li-ion battery are directly related to its components, namely, the electrodes and the electrolyte[12-14]. Present battery technology uses lithium cobalt oxide ($LiCoO_2$) and a variety of doped combinations, such as NMC ($LiNi_{0.33}Co_{0.33}Mn_{0.33}O_2$) or NCA ($LiNi_{0.8}Co_{0.15}Al_{0.05}O_2$), as the cathode[12, 13, 15, 16]. A major problem with these materials is that they contain cobalt in significant amount, which limits their use owing to toxicity, high cost, low thermal stability, and fast capacity fade at high current rates. Therefore, alternative materials are actively being investigated as possible replacements[15, 17-19]. $LiAlO_2$ is considered a potential alternative cathode material. It is also an important material in the microelectronics industry as it exhibits very small tunable lattice changes during lithium diffusion, which makes it suitable as a substrate material for epitaxial growth of III–V semiconductors like GaN[20]. The diffusion of lithium is considered as an important factor for the design, stability, and performance of $LiAlO_2$ based devices. In recent years, $LiAlO_2$ has been extensively studied, using a wide range of experimental and computational techniques[21-28], for its interesting properties as a Li-ion battery material. Phase transformations occurring in $LiAlO_2$ upon H-adsorption have been investigated[29] using combined experimental and first-principles studies. In the nuclear industry, $LiAlO_2$ is also considered as blanket material for the tritium breeder in nuclear fusion reactors [30, 31]. Classical molecular dynamics simulations suggest[32] that tritium migration is promoted and controlled by lithium diffusion. The radiation damage process in $LiAlO_2$ has also been studied[33] on an atomic scale using classical MD simulations.

Six polymorphs of $LiAlO_2$ are reported in the literature. However, the structures of only four of these (α, β, γ, and δ) are known[34]. The γ-$LiAlO_2$ is the only stable phase under ambient conditions. The structure of γ-$LiAlO_2$ (**Fig. 1,** tetragonal space group $P4_12_12$, Z=4) has been investigated using neutron and X-ray diffraction techniques[22, 35]. The structure consists of $AlO_4$ and $LiO_4$ polyhedral units sharing one edge. The γ phase is stable[36] up to 1873 K. The temperature dependence of zone-center Raman active modes has been reported in γ-$LiAlO_2$ from 78 to 873 K, which showed a considerable broadening of the phonon modes involving Li vibrations[27]. The dynamics of Li ions in a $LiAlO_2$ single crystal has been studied by NMR spectroscopy and conductivity measurements, from which a Li activation energy ~1.14 eV and a diffusion coefficient ~ $10^{-13}$ $m^2$/sec were estimated at 1000K[28]. However, the diffusion pathway for Li could not be determined from these NMR and conductivity measurements[28], but it was speculated that Li jump diffusion proceeds between tetrahedral sites via an intermediate octahedral sites[28]. Another investigation of Li diffusion based on high-temperature neutron diffraction, tracer diffusion, and conductivity spectroscopy experiments estimated the Li activation energy to be ~0.72eV and ~1.0-1.2 eV for powder and single-crystal samples, respectively[22, 23]. The authors explained the difference in barrier energy between these two forms of sample with the presence in larger concentrations of defects and vacancies in the powder sample than in the single crystal. The study in amorphous LiAlO$_2$ from room temperature and 473 K showed [37] that the lithium diffusivity obeys the Arrhenius law with an activation energy of 0.94 eV and significantly higher Li diffusion coefficient 3.6 ×10$^{-15}$ m$^2$/sec at 473K , which is three order of magnitude higher than in crystalline phase (~10$^{-18}$ m$^2$/sec) at same temperature[28]. A study on disordered γ-LiAlO$_2$ showed that the introduction of structural disorder significantly enhances the Li ion conductivity in γ-LiAlO$_2$[26].

Several new solid superionic electrolytes have been discovered in which the diffusion mechanism is found to be strongly correlated with the rotational dynamics of constituent polyhedra[38]. We earlier performed[39] extensive ab-initio calculations of phonons, high-pressure phase stability, and thermal expansion behaviour in different phases of LiAlO$_2$. We have also investigated the thermodynamic and ionic transport properties in several Li- and Na-based solid-ionic conductors using AIMD simulations and neutron scattering measurements[40-44]. Here, we focus on the diffusion mechanism of Li in various forms of LiAlO$_2$ and investigated the structural and dynamical features relevant to Li diffusion. The compound LiAlO$_2$ exhibits a simple framework structure in its γ-phase where the AlO$_4$ polyhedral units form a three-dimensional network connected via terminal oxygen, and the Li atoms occupy the octahedral voids formed between these tetrahedral units. The ionic conductivity in the perfect crystalline structure is very limited; however, the amorphous phase exhibits significantly higher ionic conductivity. So, we are asking two main questions: (i) does the activation of AlO$_4$ rotation lead to Li diffusion in the crystalline phase? (ii) how does amorphization promote Li diffusion? To address these questions, we performed AIMD simulations for both the crystalline and amorphous structures. Further, we also investigate the role of Li vacancies in ionic diffusion. Phonons are important as they dynamically modulate the structure and hence may affect the diffusion behavior. Thus, it is interesting to investigate the behaviour of phonon spectra as a function of temperature and examine their possible correlation with Li diffusion. We have measured the temperature dependence of phonon spectra in γ-LiAlO$_2$ up to 1073 K and the observed spectral changes are interpreted using AIMD simulations.

## II. EXPERIMENTAL

A polycrystalline sample of γ-LiAlO$_2$ was prepared[35] by solid-state reaction of Li$_2$CO$_3$ and gamma-Al$_2$O$_3$. Gamma-Al$_2$O$_3$ was heated at 700°C overnight prior to use. Desired amounts of Li$_2$CO$_3$ and pre-heated

gamma-Al$_2$O$_3$ in 1.01:1.00 molar ratio, were mixed thoroughly and pressed into pellets of 20 mm diameter and 10 mm height. A slight excess of Li$_2$CO$_3$ was employed to compensate for the loss of Li$_2$CO$_3$ at high temperatures. The pellets were heated at 600°C for 12 h and then crushed to powder and repelletized. These pellets were again heated at 800°C for 24 h and then the temperature was raised to 950°C and held for 24 h. The bright white pellets were crushed to powder and characterized by powder XRD. The formation of phase pure tetragonal γ-LiAlO$_2$ was confirmed by comparing the XRD data with that reported in the literature[22, 23].

The inelastic neutron scattering (INS) measurements of the phonon density of states (DOS) on the γ-phase of LiAlO$_2$ were carried out using the time-of-flight spectrometer IN4C at the Institut Laue Langevin (ILL), France. Thermal neutrons of wavelength 2.4 Å (14.2 meV) were used for the measurements, which were performed in neutron energy gain mode. We used 2 cm$^3$ of polycrystalline sample of γ-phase of LiAlO$_2$ for the measurements. The polycrystalline sample was loaded inside an 8 mm diameter cylindrical can of niobium. The data from the sample and empty niobium can were collected at several temperatures from 473 K to 1073 K. The detector bank at IN4C covered a wide range scattering angle from 10° to 110°. The data analysis was carried out in the incoherent one-phonon approximation to extract the DOS. In this approximation, the measured scattering function, $S(Q,E)$, with E and Q are the energy transfer and momentum transfer vector, respectively, is related[45-47] to the neutron-weighted phonon DOS, $g^{(n)}(E)$, as follows:

$$g^{(n)}(E) = A < \frac{e^{2W(Q)}}{Q^2} \frac{E}{n(E,T)+\frac{1}{2}\pm\frac{1}{2}} S(Q,E) > \qquad (1)$$

$$g^{(n)}(E) = B \sum_k \{\frac{4\pi b_k^2}{m_k}\} g_k(E) \qquad (2)$$

where the + or − signs correspond to the energy loss or gain of the neutrons, respectively, $n(E,T) = [\exp(E/k_B T) - 1]^{-1}$, T is temperature and k$_B$ the Boltzmann's constant. $A$ and $B$ are normalization constants and $b_k$, $m_k$, and $g_k(E)$ are, respectively, the neutron scattering length, mass, and partial density of states of the $k^{th}$ atom in the unit cell. The quantity between < > represents a suitable average over all $Q$ values at a given energy. $2W(Q)$ is the Debye-Waller factor averaged over all the atoms. The weighting factors $\frac{4\pi b_k^2}{m_k}$ for various atoms in the units of barns/amu are: 0.1974, 0.2645 and 0.0557 for Li, O and Al respectively. The values of neutron scattering lengths for various atoms can be found from Ref.[48].

## III. COMPUTATIONAL DETAILS

The AIMD simulations were performed at several temperatures (300K, 1200K, 1400K, 1800K, and 2200K) using the VASP software[49, 50]. The simulations used a 2×2×2 supercell (128 atoms) of the ambient phase structure of γ-LiAlO$_2$ and were performed for three cases: the ideal crystalline phase, a defect structure with one Li vacancy in the 128 atoms supercell, and in an approximant of the amorphous phase. Since the AIMD simulations are computationally expensive, we used a single electronic k-point at Gamma point for the total energy calculations. The calculations used the projector augmented wave (PAW) DFT formalism within generalized the gradient approximation (GGA) parameterization by Perdew, Becke, and Ernzerhof (PBE) [51-53]. A plane wave kinetic energy cut-off of 820 eV and an energy convergence criterion of $10^{-6}$ eV were used. The time step was 2 femto-second in all simulations. The diffusion is studied using long simulation runs in the NVE ensemble. To study the diffusion of Li in γ-LiAlO$_2$ near the melting point (~1873 K), we have used a ~5.7 % larger volume at 1700 K than that at 0 K to account for the thermal expansion. The volume was estimated from the experimental data[23] of thermal expansion as well as previously reported ab-initio calculations [39].

The vacancy structure was created by removing one Li atom out of the 32 Li atoms in the 2×2×2 supercell of LiAlO$_2$. To create the amorphous phase, we started with the crystalline structure of γ-LiAlO$_2$ with 128 atoms. The crystalline phase was melted at 5000 K for 10 picoseconds. Then the melted structure was quenched to 10 K, which was further used in the simulations in the amorphous phase at various temperatures from 300 K to 1400 K. We first equilibrated the structure for 10 ps to achieve the desired temperature in NVT simulations with a Nosé thermostat[54], and an NVE ensemble was subsequently used for production runs lasting up to 60 ps.

The vibrational DOS from AIMD simulations was obtained [55] via a Fourier transform of the velocity autocorrelation function. The AIMD simulations with a larger supercell would be computationally more expensive but give better averaging over the Brillouin zone. However, the calculated spectra with a 2×2×2 supercell satisfactorily reproduce the experimental neutron spectra. The simulations performed up to 60 ps give an energy resolution of ~ 0.1 meV in the calculation of the phonon DOS, exceeding the experimental energy resolution (~1 to 10 meV with ~10% of the energy transfer).

The diffusion process can be investigated in AIMD simulations by monitoring the trajectories and the time dependence of the atomic mean-squared displacement (MSD). We calculated the MSD of different atom types as a function of time at all temperatures. The time dependence of MSD is related to the isotropic diffusion coefficient [56] by the relation:

$$D = \langle u^2 \rangle / (6\tau) \quad (3)$$

where $\langle u^2 \rangle$ is the MSD at time $\tau$ is calculated using the following equation[55, 57]

$$u^2(\tau) = \frac{1}{N_{ion}(Nstep - N\tau)} \sum_{i=1}^{N_{ion}} \sum_{j=1}^{N_{step}-N_\tau} |r_i(t_j + \tau) - r_i(t_j)|^2 \quad (4)$$

Here $r_i(t_j)$ is the position of $i^{th}$ atom at $j^{th}$ time step. $N_{step}$ is total number of simulation steps and $N_{ion}$ is total number of atoms of a given atomic species in the simulation cell. $N_\tau = \tau/(\delta t)$, where $\delta t$ is step size.

We computed the self ($g_s(r,t)$) and distinct ($g_d(r,t)$) Van Hove correlation function[58] to investigate the self and correlated diffusion in various configurations of LiAlO$_2$.

$$gs(\mathbf{r}, t) = \frac{1}{N} \langle \sum_i^N \delta(\mathbf{r} + \mathbf{r}_i(0) - \mathbf{r}_i(t)) \rangle \quad (5)$$

$$gd(\mathbf{r}, t) = \frac{1}{N} \langle \sum_i^N \sum_{j>i}^N \delta(\mathbf{r} + \mathbf{r}_i(0) - \mathbf{r}_j(t)) \rangle \quad (6)$$

Here $r_i(t)$ is the atomic position of $i^{th}$ atom at time instance $t$.

The time-averaged pair-distribution function (PDF) of various pairs of atoms in crystalline, vacancy structure and amorphous phase of γ-LiAlO$_2$ has been calculated using the following relation[59]:

$$g_{IJ}(r) = \frac{n_{IJ}(r)}{\rho_J 4\pi r^2 dr} \quad (7)$$

Where $n_{IJ}(r)$ is the average number of atoms of species $J$ in a shell of width $dr$ at distance $r$ from an atom of species $I$ and $\rho_J$ is the average number density of the species $J$.

## IV. RESULTS AND DISCUSSION

**A. Temperature Dependence of Phonon Spectra**

The temperature-dependent DOS measured with INS from 473 K to 1073 K is shown in **Fig 2(a)**. The peaks at ~20, 22, 30, 35, 40, 45, 60 and 95 meV are clearly getting broadened on warming. The broadening is particularly prominent at temperatures of 873 K and above when some of peak structure is suppressed. The broadening may be due to onset of significant diffusion of Li atoms. The observed evolution of the DOS at high temperature was compared the AIMD-simulated DOS (**Figs 2 and 3**), which incorporates anharmonic effects[60]. The experimental DOS at 473 K is compared with the AIMD calculations at 300 K (**Fig 2 (b)**). The calculated DOS at 300K agrees fairly with the experimental DOS at 473 K, validating the theoretical method. Further, in order to understand the contribution of different atomic species to the INS spectra and impact of Li dynamics, we also calculated the partial DOS from AIMD trajectories (**Fig. 4**). As can be seen on this figure, the Li contribution to the DOS extends up to 80 meV, while O and Al contribute to the entire spectral range up to 110 meV. The sharp peak-like structure in the Li partial DOS at 300 K broadens significantly on warming and becomes supressed at 2200K. We also observed a significant broadening of the Al and O partial DOS, however the peak structure in the DOS is retained. The large broadening and loss of peak structure in the Li PDOS is attributed to large MSD and Li diffusion. Hence, the observed broadening and suppression of peaks in INS spectra is mainly attributed to the large MSD of Li. Further, the anharmonicity at high temperatures and lengthening of Al-O bond results in softening and broadening of the stretching modes around 95 meV in the INS spectra. The low energy Li modes ~ are found to soften upon warming. The Raman measurements[27] show that the modes at 220 cm$^{-1}$ (27.3 meV), 366 cm$^{-1}$ (45.4 meV) and 400 cm$^{-1}$ (49.6 meV) broaden on heating and their intensity weakens dramatically. The experimental observation of (**Fig. 2(a)**) of the broadening of the peaks in the phonon spectra is in qualitative agreement with these observations from Raman measurements.

The calculated partial DOS in the vacancy structure shows similar temperature trends as the crystalline phase. Interestingly, the amorphous phase does not show any peak like structure in the partial or total DOS even at 300K. However, the spectral weight shifts towards lower energy at 1400 K, especially prominently in the Li partial DOS. The main observation from the INS spectra in γ-LiAlO2 and AIMD simulations for all three phases of LiAlO2 is the significant softening and broadening of Li vibrations, which mainly occur due to large MSD and Li diffusion at elevated temperatures.

## B. Pair-Distribution Function

The environment around the lithium atoms is important to understand the pathways of Li diffusion in the material. The computed PDF for different atom pairs in crystalline γ-LiAlO$_2$ at 300 K are shown in **Fig 5**. They show well-defined peaks corresponding to Li-O, Li-Al, Al-O, Li-Li, O-O, and Al-Al bond lengths in the crystalline phase. The first peak in g(r) for Al-O arises from the bonds at 1.80 Å, while the peak corresponding to Li-O appears at 2.00 Å. The same behaviour is also observed in the vacancy structure at 300K with slightly reduced intensity of Li-O PDF which is due to a smaller overall number of Li. The PDF of various atomic pairs in the amorphous phase at 300 K only shows clear peak features for the first and second neighbours. The first peak in Al-O PDF is found to be very sharp while peaks for longer bonds get significantly broader. This reflects the limited change in the shape of the AlO$_4$ polyhedron in the amorphization process. However, the orientational and translational ordering between polyhedral units is lost. Another polyhedral unit LiO$_4$ is relatively soft and more strongly distorted during amorphization, and therefore reveals a broad peak in Li-O PDF. All other pairs of atoms also have significantly broader PDF features, which characterises amorphization in simulation.

## C. Lithium Diffusion

In the γ-LiAlO2, Li occupies the tetrahedral sites. The framework structure formed by AlO$_4$ units provides 3-d channels for Li transportation. However, the channel diameter is not homogenous, and exhibits bottlenecks formed by polyhedral corner oxygen. Our AIMD simulations show that in the ideal crystalline structure (γ-LiAlO2) at 1800 K and above, Li diffusion occurs along these channels, but it is hindered due to strong repulsive interaction between O and Li at bottleneck points. The Li ions may pass through these strongly repulsive barriers either with sufficient kinetic energy (high temperature) or upon expanding the diameter of the bottleneck (by expanding the lattice or reorienting the polyhedral units). Interestingly, by amorphization, one can achieve randomly oriented polyhedral units as well as lower density of the material, which may facilitate the diffusion process. We simulated the amorphous structure through quenched annealing as described above. The AIMD simulations of the crystalline and amorphous phases bring out the role of phonons and polyhedral reorientation on the diffusion behaviour. In the following section, we will describe the mechanism of Li diffusion in all three configurations of LiAlO$_2$.

**C.1 Perfect Crystalline LiAlO$_2$**

The calculated MSD of different atoms as a function of time in the perfect crystalline phase of LiAlO$_2$ at 300 K, 1200 K, 1600 K, 1800 K, 2000 K, and 2200 K are plotted in **Fig 6 (a)**. Below 1600 K the MSDs oscillate about their respective mean values and do not show any drift in time, corresponding to the regular behavior of a non-diffusing material. At temperatures above 1800 K, we observe a linear increase in time of Li MSD while Al and O MSDs remain constant around their mean positions. This indicates the diffusion of Li within a stable host lattice. The host lattice stability at elevated temperatures is one of the important and desired property of solid-state battery materials.

In **Fig 6(a)**, we have shown the MSD values averaged over all the atoms of each species. However, monitoring of the individual atomic trajectory and MSD provides insights into the diffusion behaviour. Hence, in **Fig 7**, we have also plotted the MSD of individual Li atoms. To explain the observed jump in individual Li MSD, we provide the Li-Li bond distances up to sixth neighbour in **Table I.** The first-neighbour distance between Li-Li is ~3.18 Å, and a jump along this bond contributes an step increase in MSD ~10.1Å$^2$ (Li has four first neighbours in LiAlO$_2$). It can be observed from **Fig 7** that most of the MSD steps in crystalline and vacancy LiAlO2 at a800K and 2200K are dominated by first neighbour jumps. The second-neighbour distance of Li-Li is ~4.26 Å (MSD~18 Å$^2$); it is interesting to note that we do not observe any direct second neighbour Li-Li jump. Hence, these observations clearly suggest that the barrier energy along pathways connecting the first neighbours is smaller and different from second neighbour pathways. The absence of second neighbour jump is due to much closer distances between Li and O along the pathways. In **Fig 7**, we also observe a higher jump in MSD values ~25 Å$^2$, which corresponds to the third-neighbour Li-Li pairs. We also notice from **Fig 7** that in the crystalline structure at 1800 K, simultaneous jumps occur for several atoms, revealing a correlated jump behavior. Isolated jumps between nearest neighbour sites in the prefect crystalline structure are not probable since there are no vacant sites. However, as we will discuss below, such independent jumps do occur in the vacancy structure and the amorphous structure.

As the temperature is increased to 2200 K, Li atoms show (**Fig. 7**) large displacements in the AIMD simulations. These MSDs match very well with the calculated jump displacements of lithium atoms between the third, fourth and fifth neighbor sites (Table I) of 25.40 Å$^2$, 28.41 Å$^2$ and 40.96 Å$^2$, respectively, which confirms that the diffusion process is governed by the jump- diffusion model[61, 62].

Further, the calculated anisotropic components of the Li MSD averaged over all Li atoms show (**Fig. 8**) that the diffusion of Li atoms at short times is easier in the tetragonal *x-y* plane than along the *z*-axis of the unit cell. The anisotropic components of individual Li MSD's are shown in **Fig. 9**. These anisotropic MSD components of individual MSD's can be better visualized with the help of calculating the Li-Li distance components in x-y plane and z- axis. From Table I, the nearest in-plane (x-y plane) jump distance is ~7.6 Å$^2$, and $\Delta z^2$ along the c-axis is 2.56 Å$^2$. However, as the simulation is continued for longer times, further jumps beyond the first neighbours take place (Table I) and the net diffusion becomes isotropic.

The AIMD calculated trajectories of selected Li atoms in the crystalline phase of γ-LiAlO$_2$ are shown in **Fig.10**, which confirms that at 1800 K there are jumps of Li atoms from one tetrahedral site to another. As the temperature is increased to 2200 K, the diffusion of Li is also through longer jumps; however, we also find an increase in the number of Li jumps from one tetrahedral site to another site. Large amplitude librations and reorientations of AlO$_4$ polyhedra help the Li diffusion process, as we discuss later.

### C.2 Vacancy in Crystalline LiAlO$_2$

Vacancies and defects are known to enhance the diffusion behaviour[63][64, 65]. Usually, in solid ionic conductors the diffusing ions hop from one occupied site to another unoccupied site. The unoccupied sites could be interstitial or atomic sites. In absence of unoccupied sites in crystals, the ions can diffuse by simultaneous jumps of two or more ions, leading to a cooperative or correlated diffusion. Also, in the case of perfectly crystalline phase, the Li diffusion is limited since it has to overcome the interaction energy of nearby host elements as well as from Coulomb interaction with the nearby occupied Li sites. Fully occupied sites along the diffusion path preclude adjusting or reorganizing the host framework to create enough room for Li migration. But the presence of vacancies usually enhances the diffusion behaviour by relaxing these constraints. We investigated the effect of Li vacancies in LiAlO$_2$, with one Li vacancy among 32 Li sites in the 2 ×2× 2 supercell of crystalline γ-LiAlO$_2$. The calculated MSD of each atoms in the vacancy structure from 300K to 1800K is shown in **Fig .6**. We did not observe Li diffusion below 1200 K within a simulation time ~20 ps. At 1200K, Li ions start diffusing in the system, while the Al and Si form a stable framework. By following individual Li trajectories, we found a jump in MSD values of ~10.1 Å$^2$, which corresponds to the minimum Li-Li distance in LiAlO$_2$. At further higher temperature (1800 K and above), many such jumps are observed, with jump lengths similar to those in the perfect crystalline structure. However, fewer simultaneous jumps occur than in the crystalline structure. The MSD at 1800 K (**Figs. 6 and 10**) is about three times larger in the vacant LiAlO$_2$ than in the perfect crystalline structure. We also analysed the anisotropic MSD of individual Li atoms at 1200K and found similar jump lengths to the crystalline phase at 1800K (**Fig. 9**).

Our AIMD simulations show a significant enhancement in diffusion from the presence of Li vacancies in crystalline $LiAlO_2$. We observed similar pathways for jump-diffusion in both the case. It seems that one can achieve the Li diffusion in the presence of vacancy in systems where the presence of vacant sites helps to reduce the dependence on the polyhedral dynamics necessary for the Li diffusion process in the stoichiometric compound. The calculated trajectories of selected Li atoms in the vacancy structure (**Fig. 10**) at 1200 K and 1800 K confirm that vacancy leads to enhancement of the Li diffusion process in comparison to that in the crystalline phase. Yet, the diffusion process in both structures occurs only through discrete Li jumps from one tetrahedral site to another.

### C.3 Amorphous $LiAlO_2$

The introduction of disorder in $\gamma$-$LiAlO_2$ is known to increase [26, 37] the Li-ion conduction. The polyhedral orientational disorder in the amorphous structure is also known to enhance [38, 66-68] the diffusion behaviour in other compounds. A large number of graphite/carbon-based materials[69-71] are known to exhibits enhanced diffusion in the amorphous form compared to their crystalline phases. Our AIMD simulation in the amorphous phase of $LiAlO_2$, strikingly, shows Li diffusion occurring at 600K, which is (**Fig. 6**) much lower than in crystalline and vacant phases. The PDF of the amorphous phase at 300K (**Fig. 5**) shows that only the first-neighbour Al-O and O-O peaks are well defined, with a broad distribution at larger bond lengths suggesting orientationally disordered polyhedral units. Interestingly, the Li-Li PDF shows a broad distribution even around the first-neighbour distance at 3.0 Å. These first-neighbour Li-Li correlations are stabilized due to misoriented $AlO_4$ polyhedral units. Further, the amorphous phase exhibits significantly smaller first-neighbour Li-Li distances, which may enhance the Li hoping probability. The Al-O PDF does not show the pronounced temperature change seen in the Li-Li PDF. We also notice that the peaks in Li-O and O-O do show significant changes with temperature, which is due to the fact that at high temperatures the randomly oriented $AlO_4$ polyhedra are kinetically activated. Unlike in the crystalline phase, the diffusion in the amorphous phase appears (**Fig. 7**) to be similar to that of a liquid. The calculated trajectories of some of the Li atoms (**Fig. 10**) at 600 K and 1400 K confirm a broad distribution of Li displacements in the diffusion process. It appears that due to the misorientation of $AlO_4$ polyhedra and the more distributed Li-Li bond distances, new low energy pathways for Li diffusion are available in the amorphous structure as compared to those available in the crystalline and vacant structure.

**D. Van Hove Pair Correlation Function**

In order to further investigate the time evolution of the diffusion process, we have computed the Van Hove self $g_s(r,t)$ and distinct $g_d(r,t)$ correlation functions[58] between Li atoms. We show $g_s$ and $g_d$ in **Fig. 11,** Left and Right panels, respectively**,** at different time intervals. The calculations shown at elevated temperatures correspond to Li diffusion occurring in all three phases. At t=0, $g_s$ is a delta function at r=0, hence it is not shown. At later times, the first peak in $g_s$ at about 1 Å corresponds to Li vibrational amplitude, and the height of this peak decreases with time inferring that few of Li ions diffuse thus limiting the magnitude of $g_s$ at farther distances. In the crystalline phase, at 5 ps, a peak develops in $g_s$ at ~3 Å, and grows in intensity with time. From 10 ps onwards, we see that another peak appears at ~5 Å. These peak positions correspond to the distances between neighbouring Li sites in the crystal structure, thus suggesting that the Li ions diffuse via discrete jumps. Similarly, in the vacancy configuration, we also see peaks at the same distances of ~3 Å and ~5 Å, but the diffusion process is faster. On the other hand, in the amorphous phase, $g_s(r,t)$ does not exhibit any sharp peaks, but rather a broad feature, which reflects that the diffusive process in this phase is not restricted to jumps between periodic sites. This indicates a more homogeneous spatial environment than in the crystalline phases. This homogeneity in space is attributed to randomly oriented polyhedral units resulting in local metastable diffusion paths.

In **Fig 11 (Right panels)**, we show the Li-Li distinct Van Hove correlation function, $g_d$, which gives the probability of finding two distinct Li atoms separated by distance r and time t. In other words, given a Li atom at r=0 and t=0, this shows how the distance-distribution of other Li atoms evolves with time t. We observe a sharp peaks at t→0 due to the crystalline structure, which correspond to various Li-Li bond distances in the static structure. The diffusion process leads to a reduced probability of finding two Li at the static bond distances; however, the lost intensity is observed in between these distances. When a random diffusion process occurs the peak intensity continually decreases. However, when two Li are diffusing in a correlated manner, the intensity of the peak corresponding to interatomic distances will slightly reduce but the peak structure will remain over a long period of time. In the case of crystalline $LiAlO_2$, we observe that the first-neighbour peak intensity does not significant decay with time, indicating the short-range correlation behaviour. The jump-diffusion in the perfect crystalline structure occurs between lattice sites without any interstitial site. This necessitates correlated diffusion as there are no vacant sites available. The vacant crystalline phase shows a mix of correlated and uncorrelated jump-diffusion. The amorphous phase exhibits both jump and continuous diffusion but not correlated jump behaviour.

### E. Polyhedral Reorientation and Lithium Diffusion

Interestingly, in the crystalline phase, all the observed MSD jumps correspond to the Li-Li neighbour distance. Hence, this infers a negligible probability for the presence of Li interstitial site in the crystalline as well as in vacant $LiAlO_2$. So, without vacancy or interstitial sites, another mechanism must enable Li diffusion. This can be better understood by monitoring the Al-O bond angle projection to the z-axis and x-axis. In **Fig. 12**, we plotted the bond angle projections of a few selected polyhedra as a function of time at different temperatures in the crystalline phase. We find that at 300 K the angles are fluctuating within a few degrees, which is common in stable crystalline materials. However, at 2200 K, the angles show large fluctuation and even polyhedral reorientation. The large vibrational amplitude of $AlO_4$ units widen the Li hopping channel and significantly reduces the barrier energy for Li hoping between two Li tetrahedral sites. Hence, this shows that the Li diffusion is correlated with larger vibrational amplitude of these polyhedral units. The correlation factor also depends on the volume of the material or density. Materials with a high volume per atom may not show such correlations since the wider diffusion channels are not affected as much by framework polyhedra dynamics.

In the case of $LiAlO_2$ with vacancies, we also observe large fluctuations in $AlO_4$ bond angles. Thus vacancies and polyhedral dynamics both contribute to enhancing the diffusion of Li. Interestingly, in the amorphous phase, we do find that bond angles show a large vibrational amplitude along with a shift of their mean values. Hence the metastable frozen $AlO_4$ units in the amorphous phase start reorienting at elevated temperature which also drags/helps Li to migrate from one site to another and enhances the Li diffusion.

We analyzed the distribution of several bond angles, specifically O-Al-O and Al-O-Al. These angles at 300 K are plotted in **Fig. 13**. The O-Al-O bond angle is related to the geometry and rigidity of the $AlO_4$ polyhedral units. We can see that in both the perfect crystalline structure and the vacancy structure, the geometry of the $AlO_4$ tetrahedra is very well maintained since the average bond angle O-Al-O is close to 109°. In the amorphous phase also, we see that the bond angle distribution peaks around 109°, but with a significant spread compared to the crystalline structure, indicating some level of distortion of the tetrahedra. The Al-O-Al bond angle, corresponding to the angle between two corner-sharing tetrahedra, shows a broad distribution in the amorphous phase, which indicates that the tetrahedra are somewhat randomly oriented.

## F. Diffusion Coefficient

We calculated the diffusion coefficient as a function of temperature and used the Arrhenius relation to obtain the activation energy for the diffusion process in various phases. Following Eq. (4), the linear fit of the slope of the Li MSD with time (**Fig. 6**) at different temperatures was used to obtain the diffusion coefficient (**Fig. 16**). To estimate the activation energy, the temperature dependence of the diffusion coefficient is fitted with an Arrhenius relation, i.e.,

$$D(T) = D_0 \exp(-E_a/k_B T) \qquad (8)$$

or equivalently,

$$\ln(D(T)) = \ln(D_0) - E_a/k_B T \qquad (9)$$

where $D_0$ is a constant factor representing diffusion coefficient at infinite temperature, $E_a$ is the activation energy, $k_B$ is the Boltzmann constant and T is temperature in K units. We find that the self-diffusion coefficient of Li-ions exceeds $10^{-10}$ m$^2$/s in the crystalline phase, vacancy structure and amorphous structure at 1800 K, 1200 K, and 600 K, respectively. One may notice that the plot of ln(D) versus 1000/T (**Fig. 14**) does not appear to strictly follow the Arrhenius behaviour, which could be due to the limited length of AIMD trajectories. However, we do expect that the order of magnitude of diffusion coefficient would not change.

The activation energy for Li diffusion in the perfect crystalline phase is found to be 1.8(6) eV, which is larger than the experimentally reported value[28] 1.14 eV from single crystal experiments. In the case of the vacancy structure, we find (**Fig 14**) that the nature of diffusion changes around 1600 K, with $E_a$ = 1.2(3) eV and 0.13(6) eV above and below 1600 K respectively. As the previous study[23] also observed, a significant difference exists in the barrier energy in powder and single crystal LiAlO2 due to different concentrations of impurities and defects in these samples. Our simulation was performed in an ideal crystalline phase without extrinsic or intrinsic defect, while in the experiments, even in single crystals, the presence intrinsic defects is unavoidable, possibly accounting for the difference in calculated and measured barrier energies. However, $E_a$ = 1.2(3) eV estimated for the vacancy structure is in fair agreement with measurements, suggesting that the presence of vacancies could significantly lower the barrier measured in experiments. In the amorphous phase, we find $E_a$ = 0.26(2) eV, which is much lower than the experimental reported value of 0.9 eV[37]. We can attribute this discrepancy to our use of a small simulation cell (128 atoms) for the amorphous structure, which can lead to larger distortions in polyhedral units, and to the limited trajectory length. However, our findings about the mechanism of Li diffusion in the amorphous phase likely would not change in larger AIMD simulations. Further, the emergence of shorter pathways along with comparatively lower barrier energies attributed to misorientation and distortion of the polyhedral units in the amorphous phase led to Li diffusion at lower temperatures. The lowering of $E_a$ in the amorphous phase compared to the crystalline phases agrees with the experimental observations, showing that disorder facilitates[26, 37] Li diffusion.

## V. CONCLUSIONS

Our AIMD simulations, combined with INS measurements, enabled us to rationalize the diffusion mechanism in LiAlO$_2$. We can understand Li diffusion in γ-LiAlO$_2$ in terms of well-defined jumps between lattice sites in the crystal, while a combination of jump-like and continuous diffusion occurs in the amorphous phase. The activation energy for the diffusion processes have been obtained in various phases. The presence of Li vacancies and disorder in the amorphous structure significantly reduces the Li diffusion onset temperature in comparison to the ideal crystalline phase. The lower onset temperature in the amorphous phase along with lower activation energy and larger diffusion coefficient occur due to availability of additional diffusion pathways around favourably oriented AlO$_4$ polyhedra as compared to those in the crystalline structure. Our temperature-dependent INS measurements and *ab-initio* molecular dynamics simulations also show that the Li diffusion is reflected in the significant broadening of the phonon spectrum.

## ACKNOWLEDGMENTS

The use of ANUPAM super-computing facility at BARC is acknowledged. SLC thanks the financial support of the Indian National Science Academy for award of an INSA Senior Scientist position.

TABLE I Calculated Li-Li distance vectors and the distances in γ-LiAlO$_2$ at 300 K up to fifth nearest neighbours. The unit cell has four Li atoms at Li0 (0.68613, 0.31387, 0.25), Li1(0.18613, 0.18613, 0.5), Li2(0.31387, 0.68613, 0.75), Li3 (0.81387, 0.81387, 0). The unit cell parameters as used in ab-initio calculations are a=b=5.3302 Å, c= 6.4016 Å. ΔX, ΔY and ΔZ are the difference in the X, Y and Z co-ordinates between two lithium atoms. d is interatomic distance between two lithium atoms.

|  | ΔX fractional | ΔY fractional | ΔZ fractional | ΔX (Å) | ΔY (Å) | ΔZ (Å) | ΔX$^2$ (Å$^2$) | ΔY$^2$ (Å$^2$) | ΔZ$^2$ (Å$^2$) | d (Å) | d$^2$ (Å$^2$) |
|---|---|---|---|---|---|---|---|---|---|---|---|
| Li1-Li0 | -0.5 | -0.1277 | 0.25 | -2.67 | -0.68 | 1.60 | 7.12 | 0.46 | 2.56 | 3.18 | 10.11 |
| Li2-Li0 | -0.3723 | 0.3723 | 0.5 | -1.98 | 1.98 | 3.2 | 3.92 | 3.92 | 10.24 | 4.26 | 18.15 |
| Li3-Li1 | 0.6277 | 0.3723 | -0.5 | 3.35 | 1.98 | -3.2 | 11.22 | 3.92 | 10.24 | 5.04 | 25.40 |
| Li0-Li0 | 0 | 1 | 0 | 0 | 5.33 | 0 | 0 | 28.4 | 0 | 5.33 | 28.41 |
| Li0-Li0 | 0 | 0 | 1.0 | 0 | 0 | 6.40 | 0 | 0 | 40.98 | 6.40 | 40.96 |

FIG 1 (Color online) (Left) The crystal structure of γ-LiAlO₂. Red, blue and violet spheres represent the oxygen, lithium and aluminium atoms respectively at their lattice sites. The polyhedral units around Li and Al are shown in blue and violet color respectively. (Right) Lithium atoms in a supercell of 2 × 2 × 2. Li0, Li, Li3 and Li3 correspond to the four lithium atoms in the unit cell of γ-LiAlO₂.

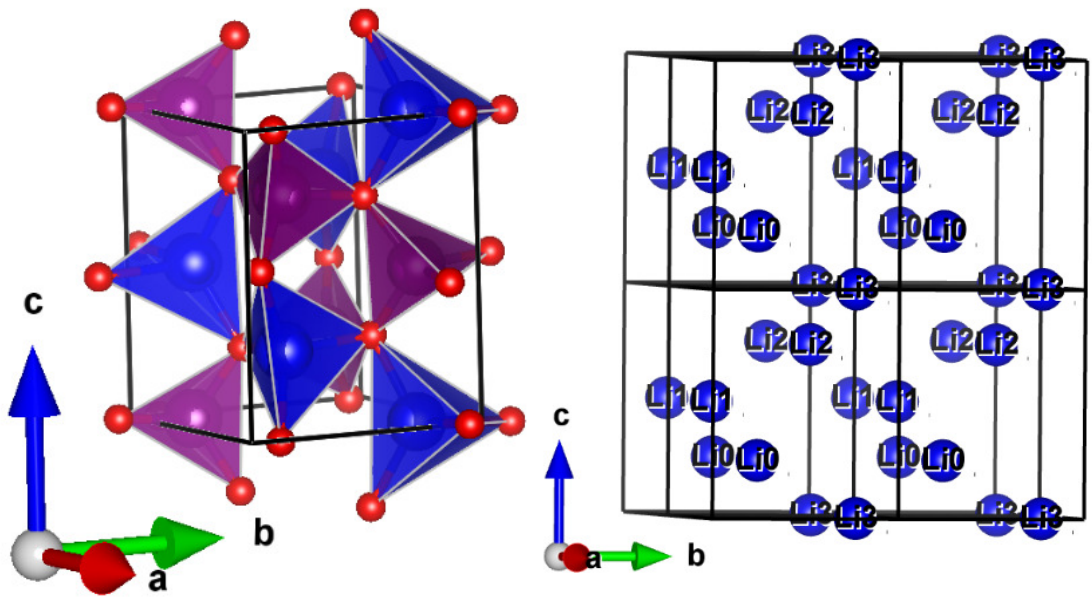

FIG. 2 (Color online) (a) The experimentally measured phonon density of states in γ-LiAlO₂ at various temperatures obtained using inelastic neutron scattering measurements. (b) Experimental (473K) and AIMD calculated (300K) neutron-weighted phonon density of states. The calculated partial contributions from various atoms are also shown.

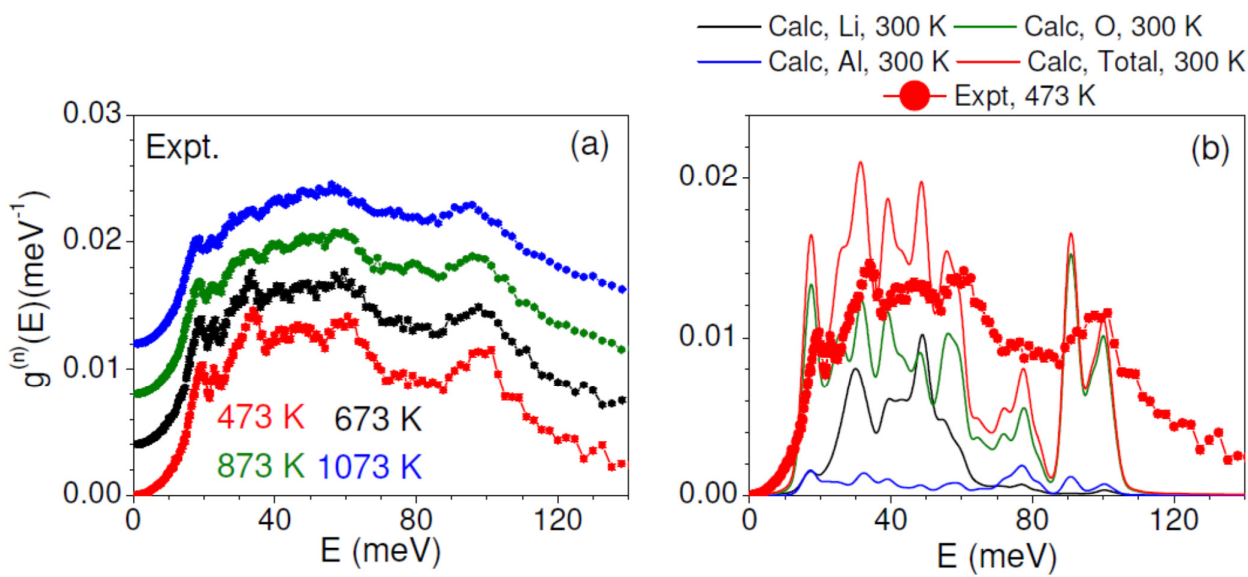

FIG. 3 (Color online) The temperature dependence of the AIMD calculated and experimental neutron-weighted phonon density of states in crystalline γ-LiAlO$_2$.

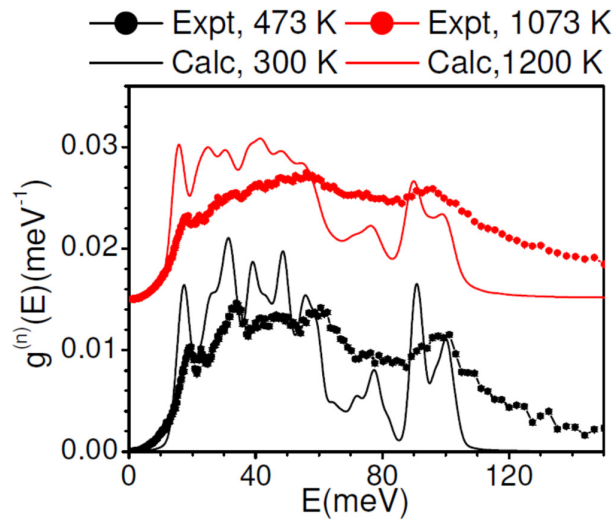

FIG 4 (Color online) The AIMD calculated partial and total phonon density of states in the perfect crystalline phase, vacancy structure and amorphous phase of γ-LiAlO$_2$.

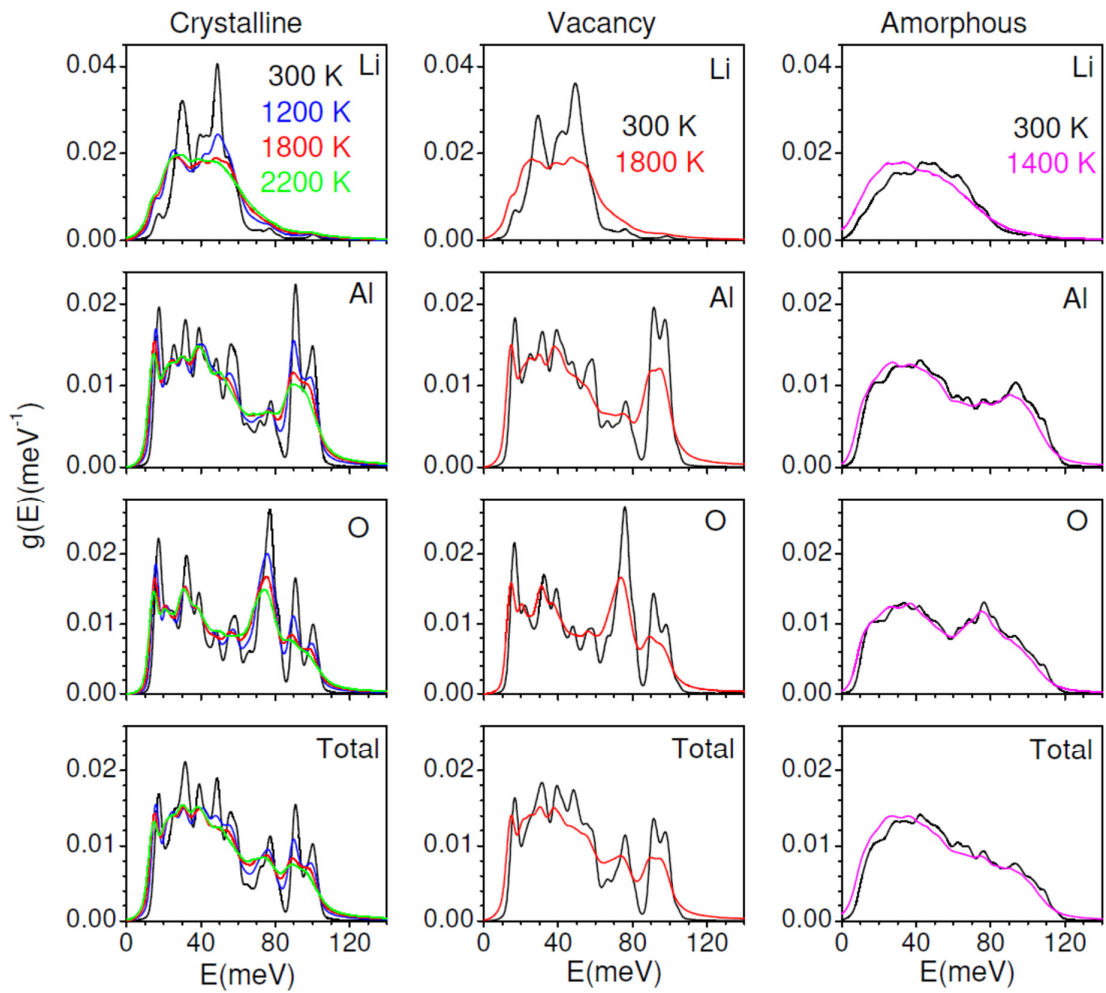

FIG 5 (Color online). The calculated pair distribution functions (PDF) of various pairs of atoms in the perfect crystalline phase, vacancy structure and amorphous phase of $LiAlO_2$. (a) shows results at 300K for all phases while (b) shows high-temperature PDF at 2200 K (crystalline), 1800 K (vacancy), and 1400 K (amorphous).

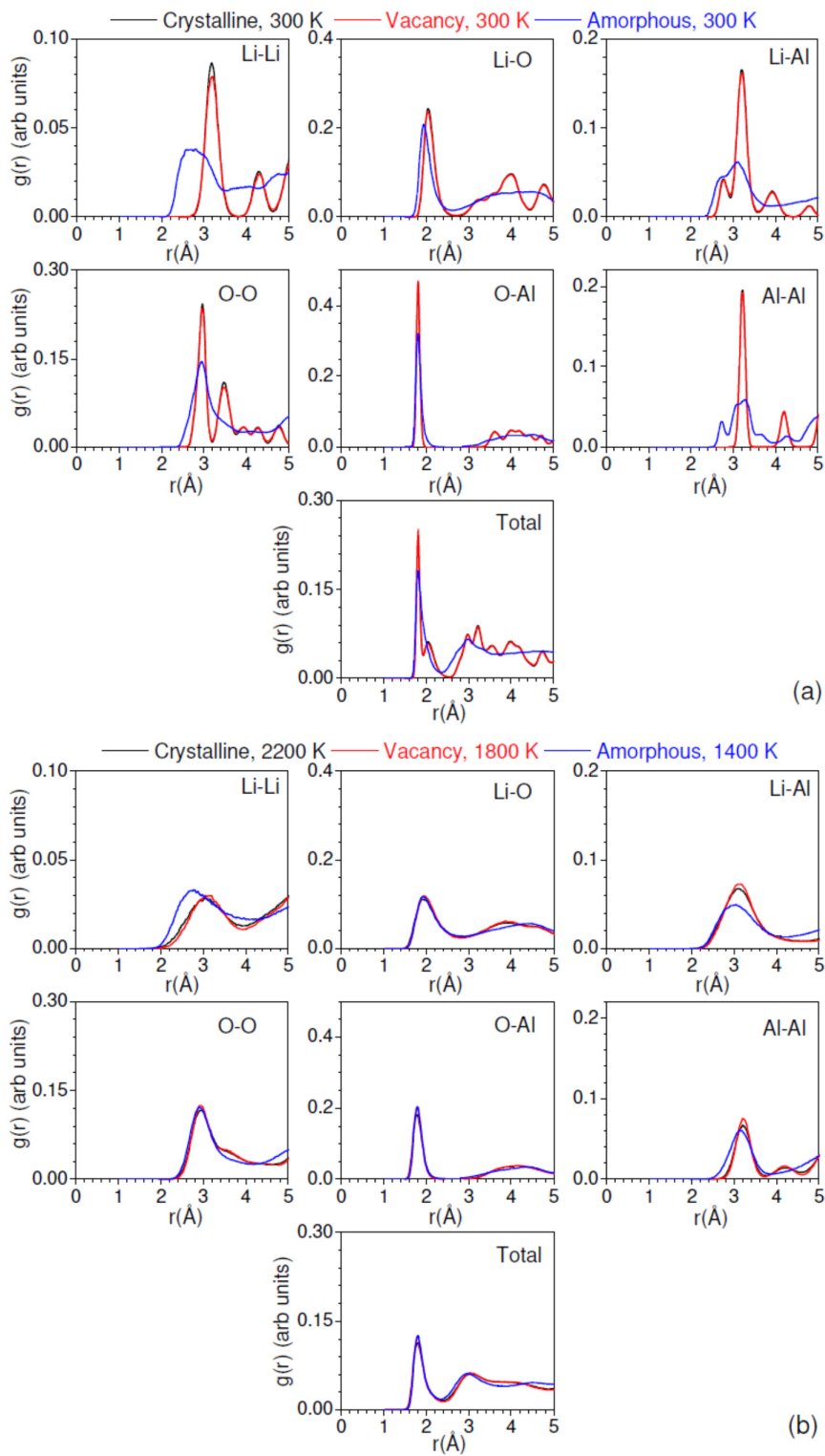

FIG 6 (Color online). Time-dependent mean-squared displacements ($<u^2> = <u_x^2> + <u_y^2> + <u_z^2>$) for each atom type in the perfect crystalline phase of $\gamma$-LiAlO$_2$, vacant LiAlO$_2$, and amorphous LiAlO$_2$ as obtained from AIMD simulations at different temperatures (indicated in each panel).

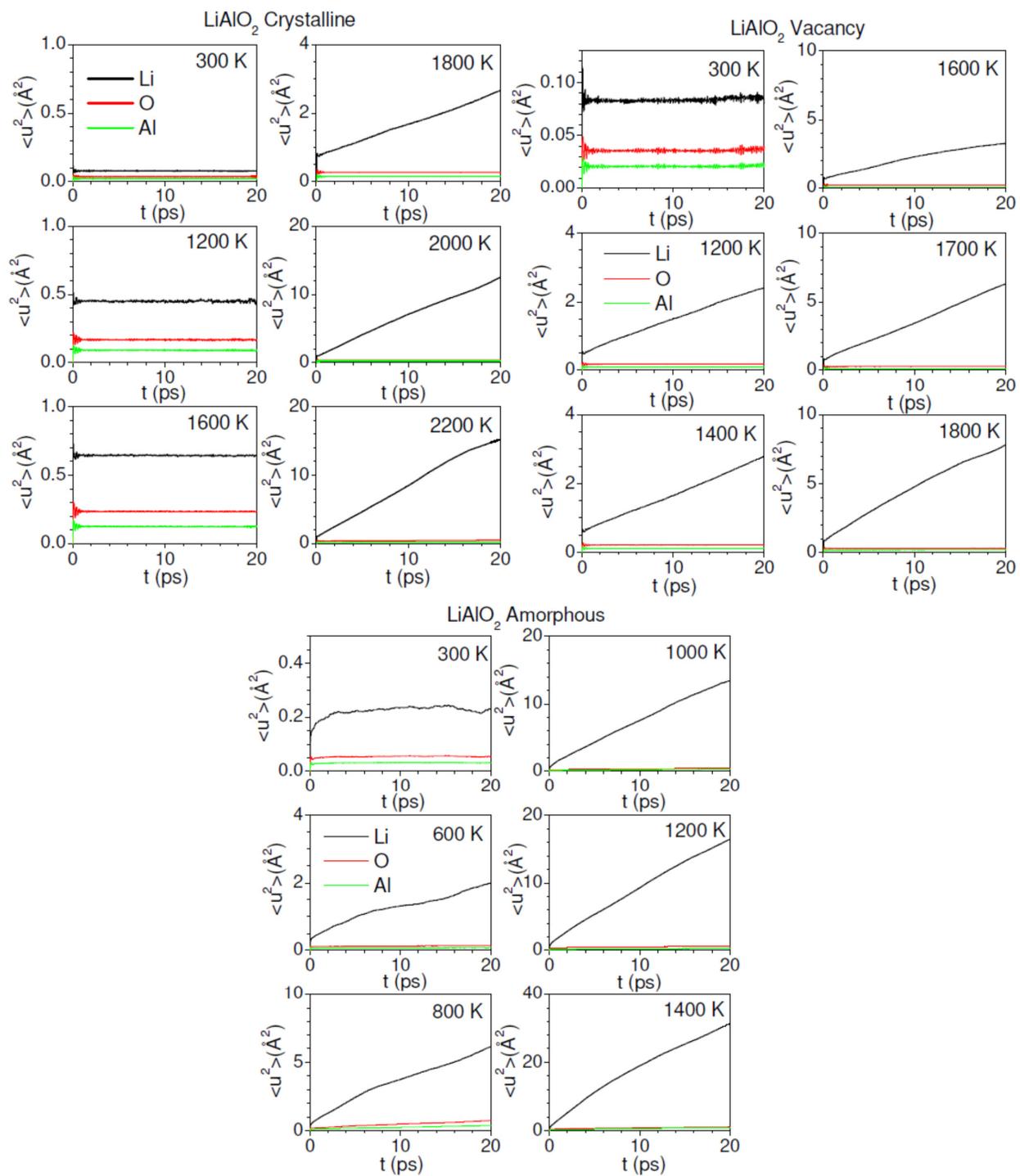

FIG 7 (Color online). The AIMD calculated squared displacement ($u^2$) of individual Li atoms as a function of time in the perfect crystalline phase, vacancy structure and amorphous phase of $LiAlO_2$ at different temperatures (indicated in each panel) The plots of various lithium atoms are shown by different colours in order to distinguish the jumps of different atoms.

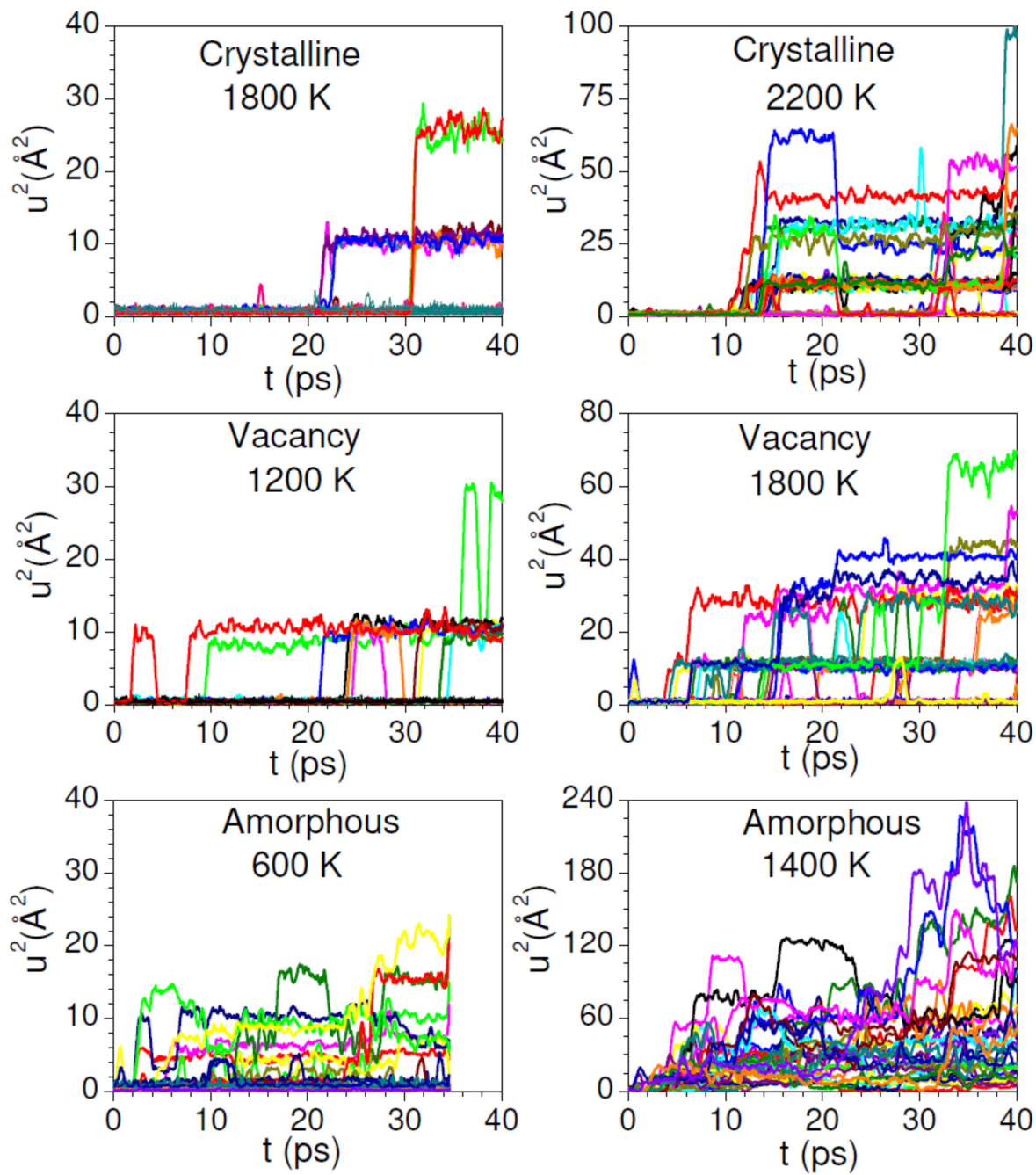

FIG 8 (Color online) The AIMD calculated anisotropic mean-squared displacement ($<u_x^2>$ and $<u_z^2>$) of Li atoms in the a-b plane and along the c-axis in the perfect crystalline phase and vacancy structure of γ-LiAlO$_2$.

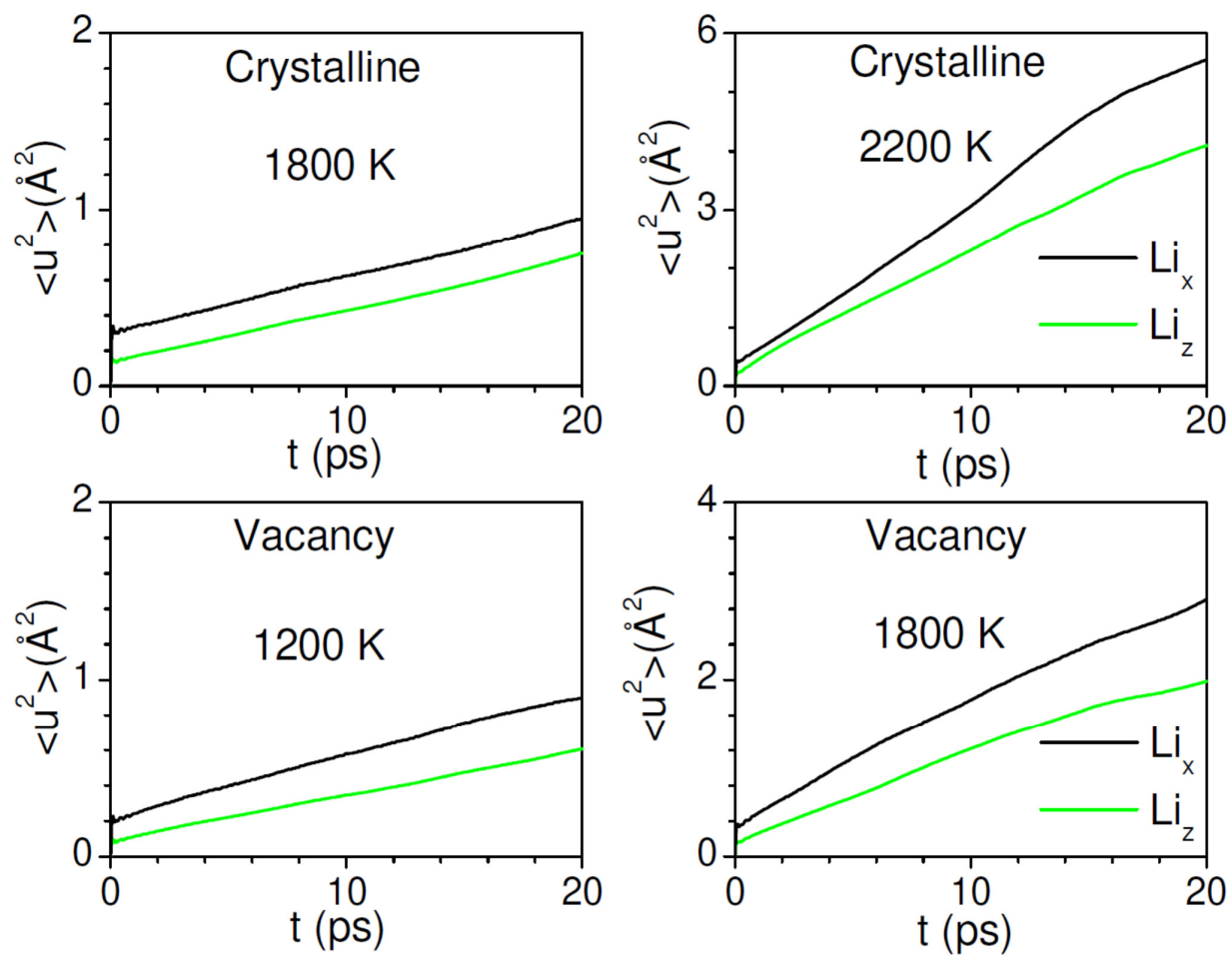

FIG 9 (Color online). The AIMD calculated squared displacements ($u^2$) of individual Li atoms as a function of time in the a-b plane and along the c-axis in the perfect crystalline phase and vacancy structure of γ-LiAlO$_2$. The plots for a given lithium atom along the x, y and z axes are shown by the same colour.

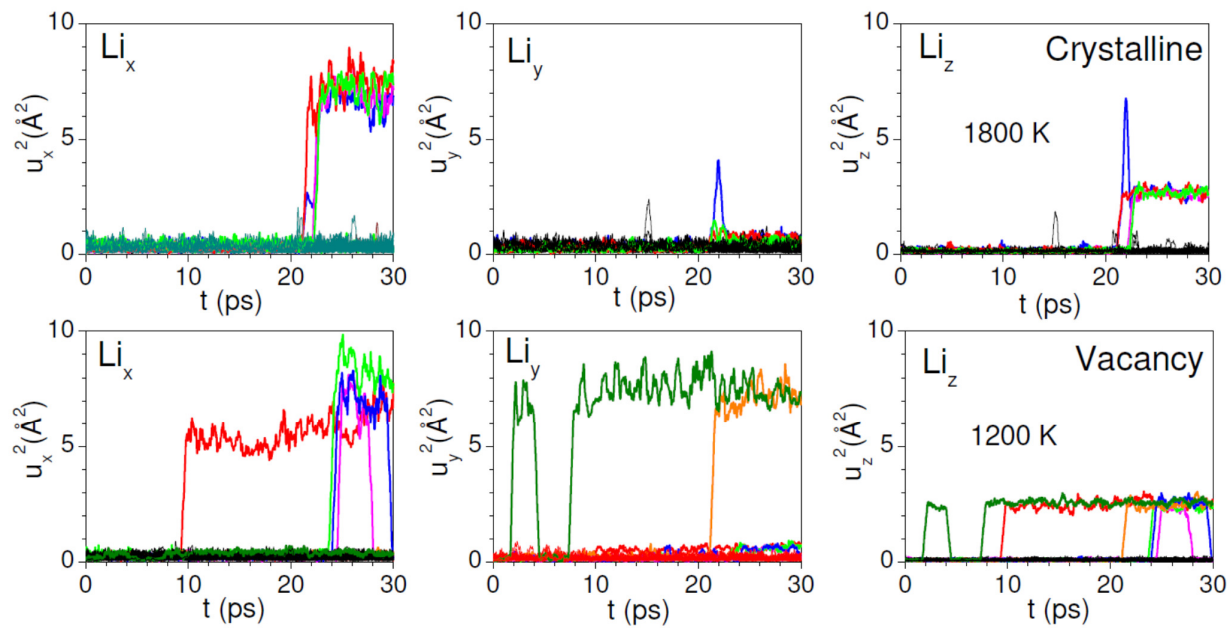

FIG 10 (Color online) Computed trajectories of selected Li atoms in the perfect crystalline phase, vacancy structure and amorphous phase of LiAlO$_2$. Red, blue and violet spheres represent oxygen, lithium and aluminium atoms respectively at their lattice sites. The time-dependent positions of the selected lithium atoms are shown by green color dots. The numbers below each frame indicate the duration of trajectory of Li and temperature of the simulation. Each figure shows the full simulation cell (2 ×2× 2 cell of γ-LiAlO$_2$).

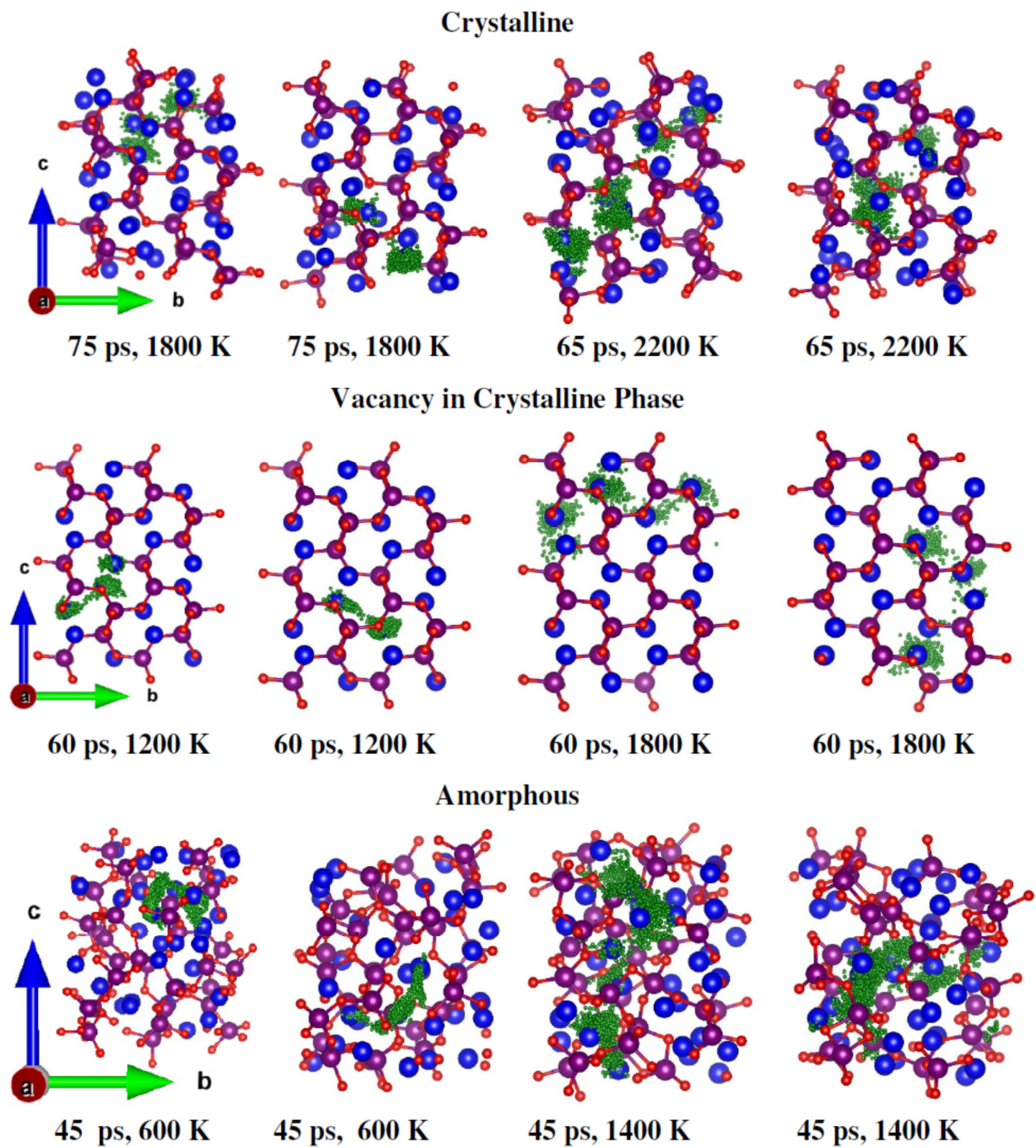

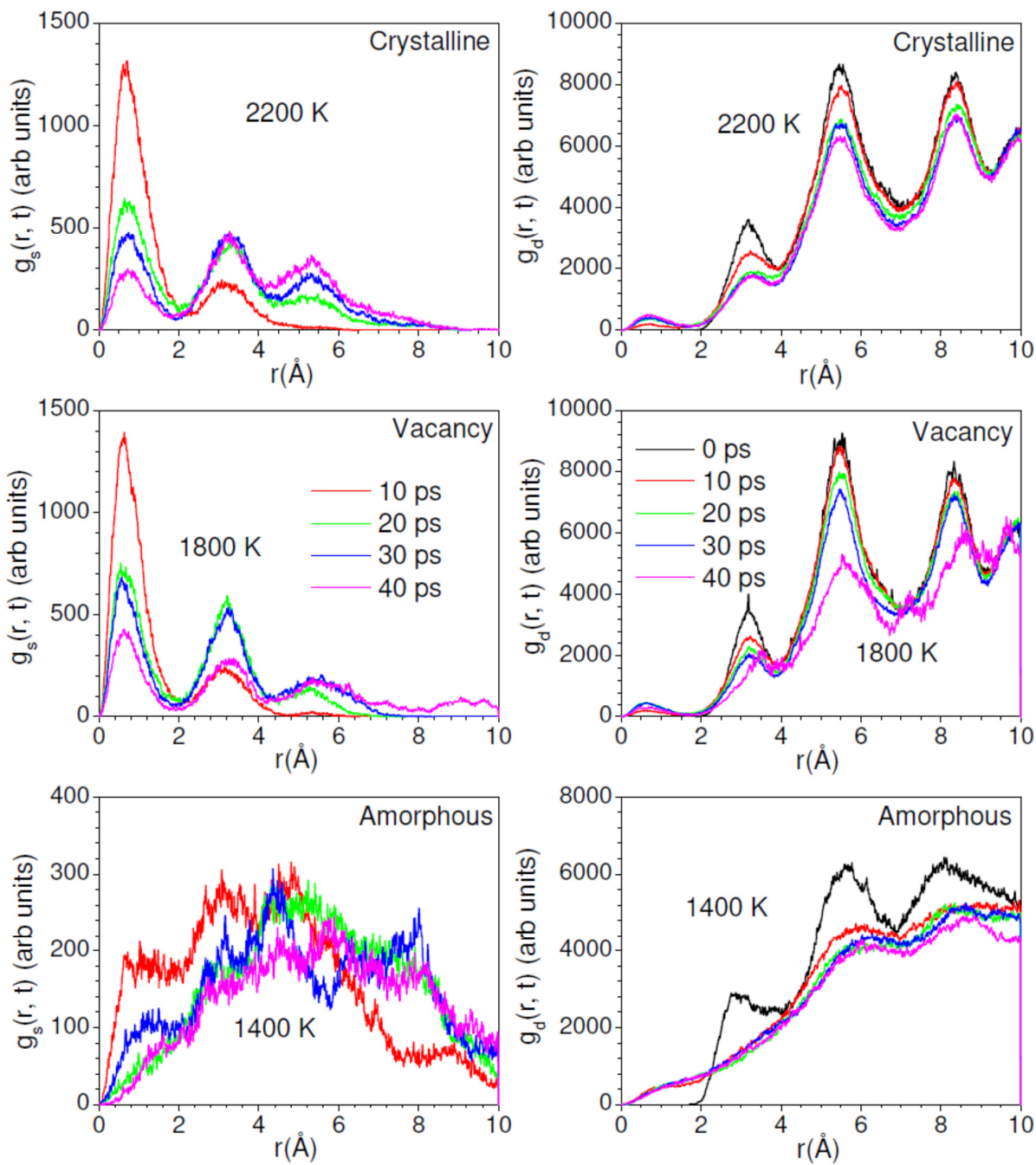

FIG 11 (Color online) The self ($g_s(r,t)$) and distinct ($g_d(r,t)$) Van-Hove correlation functions in LiAlO$_2$, computed based on AIMD, for the perfect crystalline phase, vacancy structure, and the amorphous phase.

FIG 12 (Color online) The computed angle between various Al-O bonds and crystallographic z-axis (theta) and x-axis (phi) for a representative $AlO_4$ polyhedral unit at elevated temperatures in $LiAlO_2$ for the perfect crystalline phase, vacancy structure, and the amorphous phase.

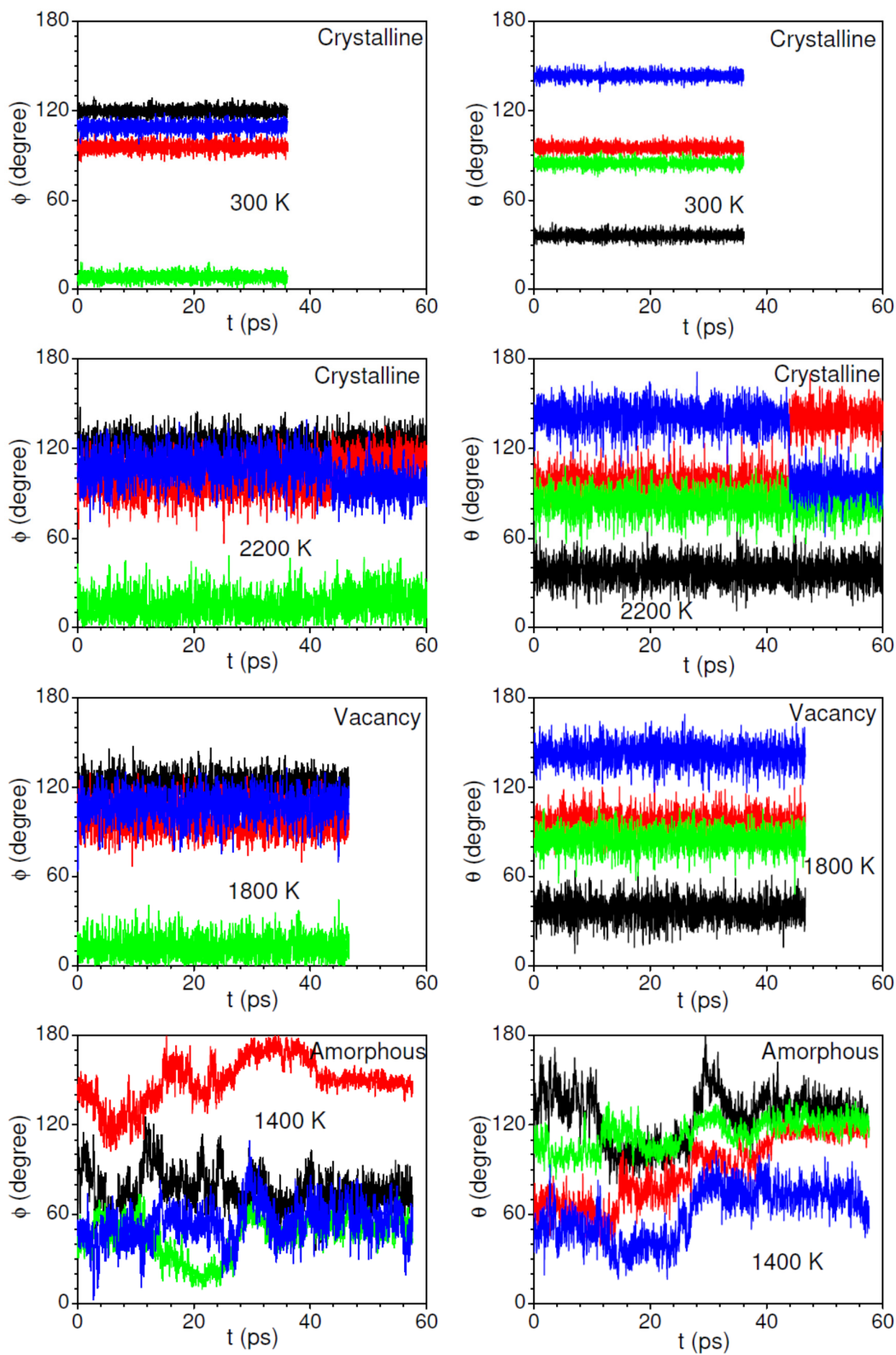

FIG 13 (Color online) Distribution of the bond-angles, O-Al-O and Al-O-Al, at 300 K.

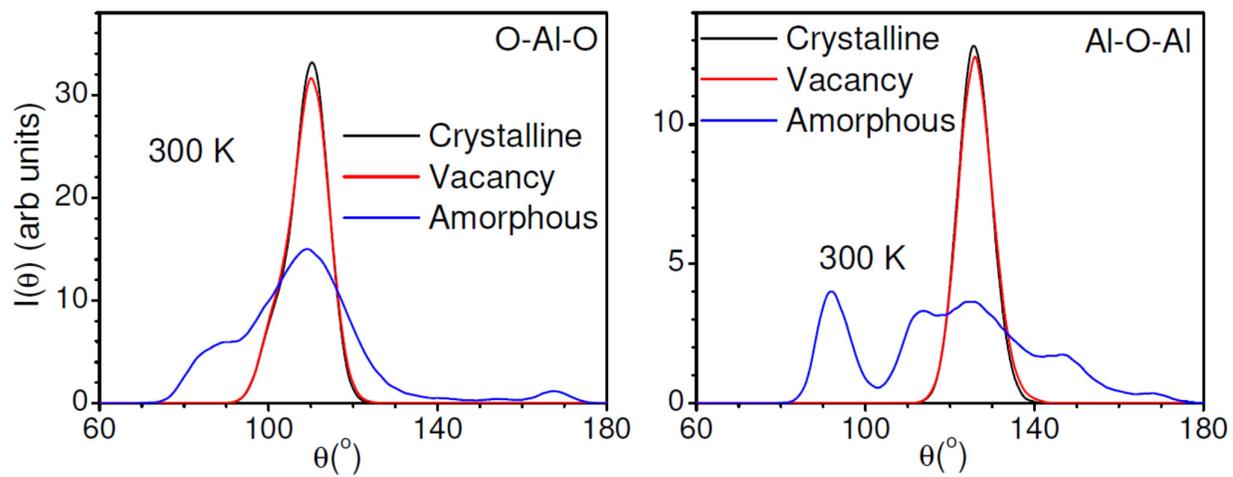

FIG 14 (Color online) The calculated diffusion coefficients and activation energy barriers in the perfect crystalline phase, vacancy structure and amorphous phase of LiAlO$_2$ using AIMD simulations.

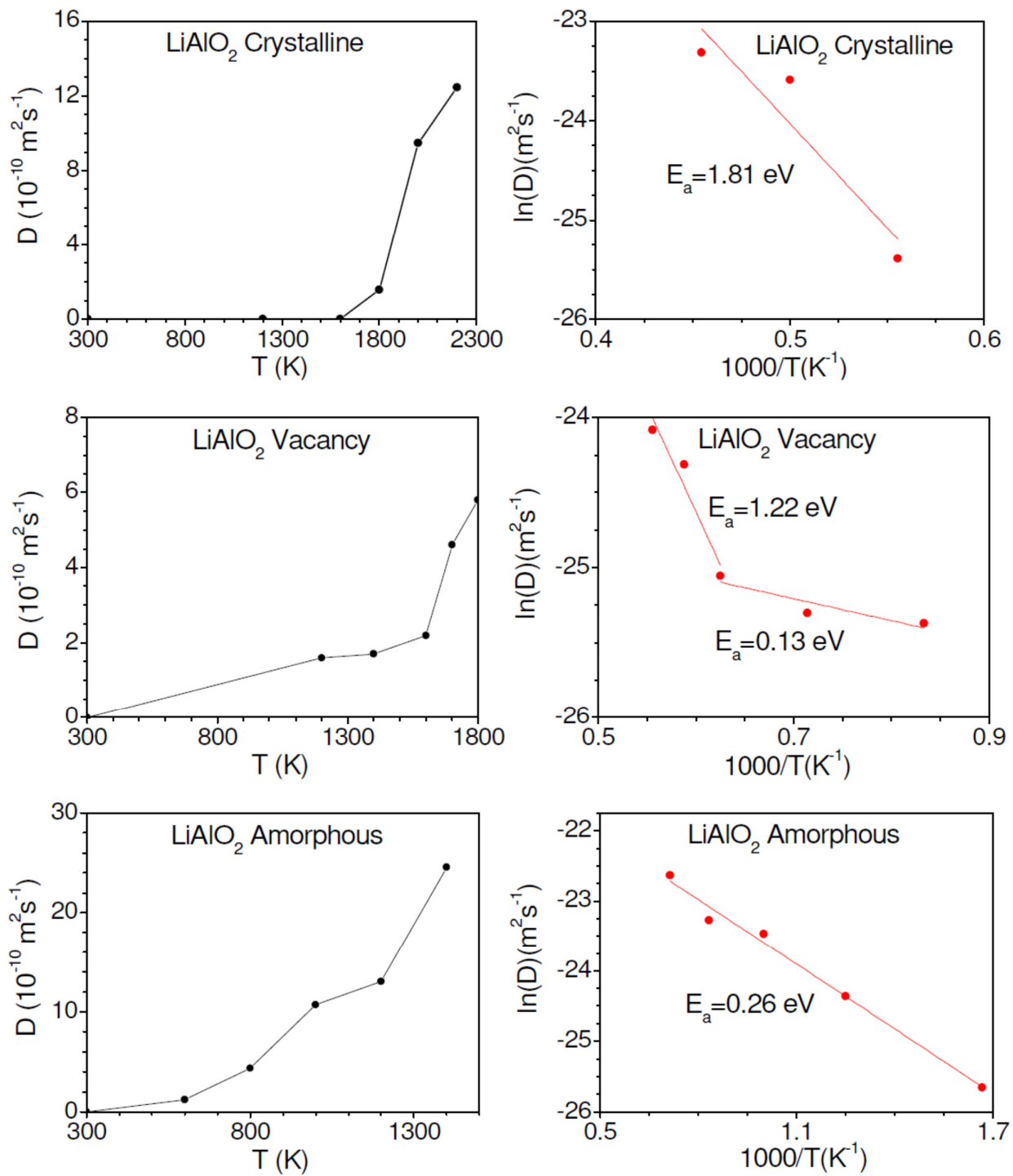